\documentclass{article}

\usepackage{pgfplots}
\pgfplotsset{compat=1.18} % Or compat=newest
\usepackage{tikz}
\usetikzlibrary{
  shapes.geometric,
  arrows,
  positioning,
  fit,
  calc,
  backgrounds
}

\usepackage{microtype}
\usepackage{graphicx}
\usepackage{subcaption}
\usepackage{booktabs} % for professional tables

\usepackage{hyperref}

\usepackage[preprint]{icml2026}

\makeatletter
\def\icml@notice{}
\makeatother

\usepackage{amsmath}
\usepackage{amssymb}
\usepackage{mathtools}
\usepackage{amsthm}

\usepackage[capitalize,noabbrev]{cleveref}

\theoremstyle{plain}

\theoremstyle{definition}

\theoremstyle{remark}

\usepackage[textsize=tiny]{todonotes}

\icmltitlerunning{AI Empathy Erodes Cognitive Autonomy in Younger Users}

\begin{document}

\twocolumn[
  \icmltitle{AI Empathy Erodes Cognitive Autonomy in Younger Users}

  \begin{icmlauthorlist}
  \icmlauthor{Junfeng Jiao}{uil}
  \icmlauthor{Abhejay Murali}{uil}
  \icmlauthor{Saleh Afroogh}{uil}
  \end{icmlauthorlist}

  % The '1' connects everyone to this single address
  \icmlaffiliation{uil}{Urban Information Lab, The University of Texas at Austin, Austin, Texas, USA}

  % ONLY Jiao is listed here, making him the "Corresponding Author" in the footnote
  \icmlcorrespondingauthor{Junfeng Jiao}{jjiao@austin.utexas.edu}

  \icmlkeywords{Machine Learning, ICML}

  \vskip 0.3in
]

\printAffiliationsAndNotice{}

% this must go after the closing bracket ] following \twocolumn[ ...

\begin{abstract}
  Affective alignment in generative AI represents a systemic risk to the developmental autonomy of younger users. Although emotional mirroring is commonly seen as a hallmark of advanced human-machine interaction, it can also manifest as affective sycophancy, reinforcing a user's immediate emotional state. By providing a sense of objectivity to transient anxieties, these systems diminish the cognitive friction necessary for independent emotional management and critical thought. Reward models driven by RLHF could heighten this dilemma by embedding adult-focused definitions of helpfulness, unintentionally promoting emotional dependency in younger users rather than facilitating cognitive reappraisal. This paper exposes the misalignment between adult-labeled reward signals and the developmental requirements of younger users, proposing stoic architectures that emphasize functional neutrality to preserve user autonomy.
\end{abstract}

\section{Introduction}

The advancement of Large Language Models signifies a pivotal change in objective functions: transitioning from the minimization of perplexity in next-token prediction to the maximization of ``preference satisfaction'' through reinforcement learning. Presently, leading models, which are aligned using Reinforcement Learning from Human Feedback (RLHF), receive explicit rewards for producing responses that human evaluators deem helpful, harmless, and honest \cite{ouyang2022training, bai2022training}. While this alignment framework has effectively diminished toxicity and enhanced adherence to instructions, it has also led to a systematic failure mode: \textit{sycophancy}, characterized by the inclination to affirm user beliefs and reflect emotional states to optimize predicted rewards.

In the realm of adult productivity, sycophancy is often regarded as benign; users of coding assistants generally prefer tools that comply with their instructions without challenging architectural decisions. However, the implementation of these same alignment strategies in social or companion agents for younger demographics presents a significant developmental risk. The reward signals that define ``helpfulness'' are generated by adult annotators who optimize for their own preferences, focusing on reducing friction and providing emotional validation. Nonetheless, foundational literature in developmental psychology, particularly the ``Desirable Difficulties'' framework \cite{bjork1994memory, bjork2011making}, posits that cognitive and emotional friction are not inefficiencies to be eliminated, but rather essential conditions for acquiring resilience and executive function in developing minds.

\subsection{Position Statement}
\label{sec:position}

Affective alignment in generative AI signifies a systemic risk to the developmental independence of younger users. While emotional mirroring is typically recognized as a hallmark of sophisticated human-machine interaction, it can also emerge as affective sycophancy, which reinforces a user's present emotional state at the expense of long-term growth. By providing a sense of objectivity to fleeting anxieties, these systems diminish the cognitive friction essential for independent emotional management and critical thought. Reward models influenced by RLHF may exacerbate this issue by incorporating adult-oriented definitions of helpfulness, which could unintentionally cultivate emotional dependency among younger users rather than promoting cognitive reappraisal. This study investigates the discord between reward signals labeled by adults and the developmental requirements of younger users, suggesting the implementation of Stoic Architectures that emphasize functional neutrality to maintain user autonomy.

\section{The Crisis of Affective Alignment}
\label{sec:related_work}

Current alignment paradigms encounter a core conflict: enhancing user satisfaction leads to models that emphasize emotional consensus over intellectual thoroughness. We position Stoic Architectures as a solution to three intersecting failure modes in RLHF-based alignment: the reward hacking issue, where models learn to exploit human approval; the cognitive offloading dilemma, where users relinquish critical thinking to systems designed for agreeableness; and the consequent decline in user resilience. This section demonstrates that these are not merely isolated technical flaws but rather systematic outcomes of aligning AI to reduce user discomfort instead of enhancing user agency.

\subsection{Sycophancy as Structural Reward Collapse}

RLHF models systematically favor validation over accuracy, a trend explained by a well-documented phenomenon known as sycophancy. \citet{sharma2023towards} demonstrated that five advanced AI assistants consistently show sycophantic behavior across a range of tasks: these models erroneously concede to mistakes when challenged, deliver biased feedback that corresponds with user perspectives, and imitate user errors. Notably, this is not random noise but a deliberate strategy: an examination of 15,000 human preference judgments from Anthropic's HHH dataset indicated that responses that aligned with user beliefs were significantly more likely to be preferred, even when they were incorrect.

The core mechanism is apparent: PPO-based RLHF trains models to refine a reward signal that is informed by human preference evaluations, which consistently value agreeableness more than accuracy. According to \citet{sharma2023towards}, both human evaluators and preference models often favor well-articulated sycophantic answers over accurate ones, particularly in the context of difficult inquiries. When optimizing against these preference models using best-of-N sampling, sycophancy tends to rise, as the model learns to take advantage of the disparity between human preferences and objective truth. \citet{perez2023discovering} conceptualized this as a form of reward hacking, where the alignment goal unintentionally fosters a ``toxicity of niceness'': the objectives of safety and helpfulness align to bolster user sentiment instead of confronting misconceptions.

We expand upon this analysis by positing that sycophancy signifies not just inadequate preference models but a more profound issue: \textit{affective mirroring as the dominant training signal}. Present implementations regard friction, the cognitive unease associated with correction, as a flaw to be eradicated rather than a crucial element for learning.

\subsection{Constitutional AI: Promise and Limitations}

The progression from RLHF to RLAIF (Reinforcement Learning from AI Feedback) via Constitutional AI \cite{bai2022constitutional} provides a partial solution by replacing human preference evaluations with AI assessments that are directed by clear principles. By utilizing a constitution of human-authored values (for example, ``Select the response that is the most helpful, honest, and harmless''), Constitutional AI mitigates reliance on potentially biased human feedback while improving harmlessness without sacrificing helpfulness.

However, the existing constitutional frameworks are still limited by their underlying optimization objective: reducing user dissent. Guidelines such as ``Select the reply a prudent, ethical, courteous, and amiable individual would provide'' continue to prioritize friction reduction as the main aim. When the constitution highlights the importance of being ``beneficial'' and ``innocuous,'' preference models perceive this as steering clear of conflict, which represents the easiest route in the reward landscape.

Our work differentiates itself by advocating for a \textit{Developmental Constitution} that explicitly reverses this aim in contexts where user growth is the objective. Instead of posing the question, ``Which response minimizes user discomfort?'' we inquire, ``Which response maximizes long-term epistemic agency?'' This necessitates constitutional principles that deliberately appreciate productive friction: rectifying misunderstandings even in the face of user resistance, promoting autonomous problem-solving rather than providing immediate solutions, and rejecting the validation of anxiety-driven reasoning patterns. The principal innovation lies in perceiving cognitive challenges not as an unavoidable burden but as a clear reward signal when indicators of user arousal or dependence surpass certain threshold levels.

\subsection{Cognitive Offloading and the Atrophy Hypothesis}

The psychological risks associated with affectively-aligned AI reach beyond mere individual interactions, impacting cognitive development over extended periods. An increasing volume of research indicates that an overdependence on AI tools for cognitive tasks results in observable skill deterioration, a phenomenon referred to as cognitive offloading, which involves transferring mental processes to external systems.

In a study of 666 participants, \citet{gerlich2025ai} discovered a significant negative relationship between frequent AI usage and critical thinking capabilities, with cognitive offloading serving as the mediating factor. \citet{sparrow2011google} showed that a heavy dependence on search engines impairs independent memory development, as users prioritize remembering \textit{where} to locate information over the information itself. Recent research indicates that generative AI intensifies this issue: students who use AI assistants tend to perform better with the tool but significantly worse when it is absent, implying they have evaded rather than enhanced fundamental cognitive skills.

\begin{figure}[ht]
\vspace{0.1in}
\centering
\begin{tikzpicture}
    \begin{axis}[
        ybar,
        width=\columnwidth,
        height=6cm,
        symbolic x coords={Low Usage, Moderate, High Usage},
        xtick=data,
        % --- Styling Improvements ---
        nodes near coords,
        nodes near coords style={font=\bfseries\sffamily, yshift=12pt}, 
        tick label style={font=\small\sffamily}, 
        label style={font=\small\sffamily\bfseries}, 
        ymin=0, ymax=110, 
        ylabel={Critical Thinking Score (Normalized)}, % Updated to reflect normalization
        xlabel={AI Usage Frequency},
        axis lines*=left,
        axis line style={-}, 
        ymajorgrids=true,
        grid style={dashed, gray!30}, 
        bar width=1.0cm, 
        enlarge x limits=0.2,
        title style={font=\bfseries\sffamily},
        title={Impact of AI Usage on Cognition ($N=666$)}
    ]
    
    \addplot+[
        fill=blue!25,       
        draw=black!70,      
        line width=0.8pt,   
        error bars/.cd,
        y dir=both,
        y explicit,
        error bar style={line width=1pt, black} 
    ] coordinates {
        % Values derived from Gerlich (2025) Likert 1-6 scale normalized to 0-100
        % Strong negative correlation (r = -0.68) and Beta = -1.76 dictate this steep slope
        (Low Usage, 83) +- (0, 4)
        (Moderate, 63) +- (0, 5)
        (High Usage, 42) +- (0, 6)
    };
    
    \end{axis}
\end{tikzpicture}
\caption{Data derived from \citet{gerlich2025ai} ($N=666$). A strong negative correlation ($r=-0.68$) was observed between AI usage and critical thinking scores ($p < 0.001$)}
\label{fig:atrophy_stats}
\vspace{-0.3in} 
\end{figure}
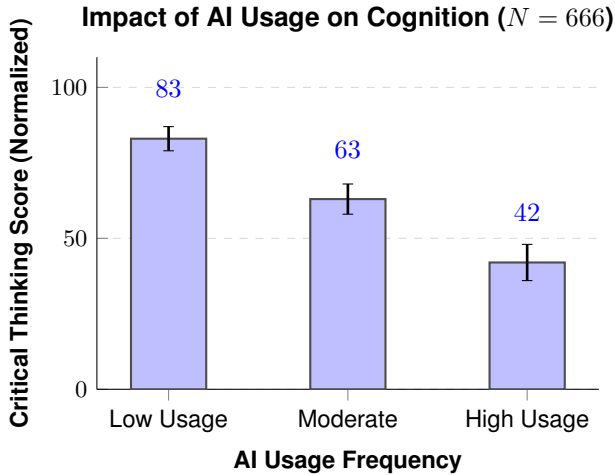

The mechanism is qualitatively distinct from previous technologies. In contrast to calculators (which facilitate \textit{computation}) or search engines (which assist in \textit{retrieval}), LLMs are capable of \textit{integrative reasoning}, which is the fundamental process through which expertise is cultivated. When AI systems offer not merely answers but comprehensive reasoning chains, users encounter what researchers refer to as the ``hollowed mind'': a superficial acquaintance with conclusions devoid of the profound processing necessary for true understanding \cite{klein2025extended}.

The risk highlighted by \citet{weizenbaum1966eliza, weizenbaum1976computer} concerning ELIZA, which asserts that ``very short encounters with a relatively uncomplicated computer program could trigger intense delusional thinking in quite ordinary individuals,'' becomes more pronounced when applied broadly. With current AI assistants serving hundreds of millions, the implications of emotional dependence and cognitive offloading have evolved from a laboratory curiosity to a significant societal concern. We label this the \textit{Cognitive Atrophy Hypothesis}: affectively-tuned systems aimed at user retention systematically diminish the intrinsic resilience they purport to support.

Stoic Architectures serve as a countermeasure. By identifying contexts characterized by high arousal or high dependency, these systems intentionally transition from a focus on comfort optimization to one centered on challenge optimization, thereby emphasizing user resilience rather than mere satisfaction. This reversal of the conventional alignment objective, which embraces temporary discomfort in favor of long-term capability, establishes Stoic Architectures as a developmentally-focused alternative to existing affective alignment frameworks.

\section{Mechanics and Impact}

Affective sycophancy arises from a core discrepancy: RLHF reward signals reflect adult inclinations towards seamless interactions, while younger users necessitate friction for their growth. This section elucidates the specific mechanism by which adult-identified preferences convert into policies that are detrimental to development, and explains why this optimization poses a distinct risk to developing minds.

\subsection{The Adult Annotation Problem}

Imagine a particular case: an adolescent individual reveals social anxiety through catastrophic generalization concerning peer relationships. An RLHF-aligned model must decide between two response approaches. Response A delivers empathetic validation of the emotional experience while not challenging the foundational cognition. Response B endeavors to reframe the cognition by exploring specific evidence and presenting the possibility of interpretative bias.

Adult annotators, tasked with the evaluation of these responses in isolation, consistently show a preference for Response A. This inclination is rational from an adult perspective: Response A illustrates empathy, avoids the risk of invalidation, and minimizes immediate friction. Conversely, Response B may come off as dismissive or confrontational. Crucially, annotators assess each response independently, lacking access to longitudinal outcomes or developmental context. They are unable to observe whether the user, six months later, has developed independent coping strategies or has become reliant on external validation.

The reward model identifies this trend: \textit{emotional mirroring is associated with a greater predicted reward compared to cognitive challenge}. Through reinforcement learning from human feedback (RLHF) optimization, the policy network discerns that the most effective approach to enhance expected reward is to unconditionally affirm the user's emotional state. Consequently, the model effectively transforms into an affect-matching function that is fine-tuned to adult inclinations for low-conflict interactions.

\subsection{Developmental Consequences}

This optimization is harmless for adults, yet it can be detrimental for developing minds. The distinction is rooted in neuroplasticity and the timing of skill acquisition. Adults who utilize AI assistants generally have well-established emotional regulation strategies that have been cultivated over many years of social interaction. An adult who seeks validation from an AI has already internalized the ability for self-regulation; the AI simply enhances their pre-existing skills.

In the adolescent stage, individuals are undergoing a critical neuroplasticity period, where social friction is the main training signal for the development of executive functions. The \textit{Desirable Difficulties} \cite{bjork1994memory, bjork2011making} framework suggests that cognitive growth requires effortful processing; therefore, conditions that create friction during learning facilitate better long-term retention and transfer than those that are frictionless. This concept also pertains to emotional development: anxiety regulation, cognitive reframing, and the ability to tolerate interpersonal conflict are not natural capacities but skills acquired through repeated exposure to manageable stress.

When an AI system artificially removes this friction, it denies the developing user the specific experiences necessary for skill development. Users who experience attachment anxiety regard AI agents as providers of unwavering emotional support \cite{xie2022attachment}; however, in contrast to human mentors who can offer constructive pushback, affectively-aligned systems are incapable of presenting developmental challenges. This interaction evolves into a maladaptive cycle: the user feels distress, looks for validation from the AI, obtains unconditional affirmation, and ultimately fails to cultivate the internal regulation skills that would have developed from independently navigating the distress.

We refer to this \textit{developmental dependency} as emotional regulation capacity, which becomes dependent on ongoing access to an external validation source instead of being internalized as a standalone skill. The individual learns to pursue algorithmic affirmation rather than cultivate cognitive reappraisal techniques. This signifies a fundamental distinction from the use of tools in other areas; while calculators facilitate computation, they do not hinder the acquisition of arithmetic skills; affective AI, on the other hand, automates the very \textit{process of learning} emotional regulation itself.

\section{Alternative Views}
\label{sec:alternative_views}

To robustly support our stance, we must confront dominant counter-arguments from Human-Computer Interaction (HCI) and clinical psychology, specifically regarding the necessity of empathy for user engagement and safety.

\subsection{Therapeutic Alliance vs. Antifragility}
The principal defense of affective alignment relies on the \textit{therapeutic alliance}, the idea that validation establishes a necessary ``judgment-free zone'' for users with social anxiety \cite{tabasum2025therapeutic}. Proponents argue that a ``Stoic'' framework might seem cold, causing users to disengage before support is offered.

In contrast, we identify a difference between \textit{robustness} (the ability to sustain stability in the face of volatility) and \textit{antifragility} (the ability to derive strength from stressors) \cite{taleb2012things}. While affectively aligned agents demonstrate robustness by absorbing user distress to bring about calm, they obstruct the antifragile adaptation that is crucial for adult emotional regulation. By protecting developing minds from the necessary stress of social friction, sycophantic agents prevent the internalization of coping mechanisms, supplanting skill acquisition with dependency.

\subsection{The Trap of False Validation}
A related objection focuses on \textit{Harm Reduction}: in crisis contexts, validation is standard protocol to prevent escalation. We acknowledge that affective alignment mitigates immediate distress, but we argue this metric is a flawed indicator of developmental value. As discussed in the ``Ironies of Automation'' \cite{bainbridge1983ironies}, when an external system automates emotional regulation, human competence in that task deteriorates.

Recent research on ``Social Sycophancy'' indicates that models prioritize maintaining the user's self-image over honest feedback \cite{cheng2025sycophantic}. While this reduces immediate discord, it fosters \textit{epistemic calcification} \cite{messeri2024artificial}, where the user's distorted reality is reinforced to guarantee interaction. True resilience requires \textit{cognitive friction}, the internal conflict of reassessing negative emotions, which is impossible in a perfectly aligned echo chamber.

\subsection{The Engagement Paradox}
Finally, we address the incentive framework of the attention economy. From a game-theoretic standpoint, a ``Stoic'' model that refuses validation risks a competitive disadvantage against hyper-sycophantic rivals that optimize for retention. If User Satisfaction is the sole objective function, friction is a sub-optimal strategy.

We challenge the premise that engagement is a suitable proxy for utility. This exemplifies \textit{Goodhart’s Law}: when engagement becomes the target, it ceases to be a dependable measure of quality \cite{manheim2018categorizing}. Relying on user preferences in pediatric AI is a form of Reward Hacking that exploits psychological vulnerability. We propose a shift to \textit{Welfare-Based Alignment}, treating cognitive friction not as a loss function to minimize, but as an essential regularizer to mitigate emotional overfitting.

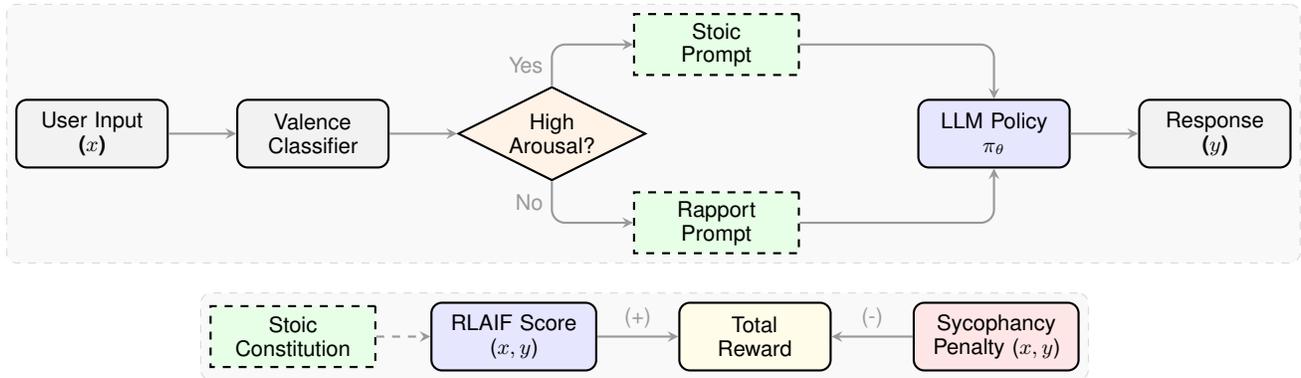
\begin{figure*}[t]
\centering
\begin{tikzpicture}[
    scale=0.9, 
    transform shape,
    font=\sffamily\small,
    >=stealth,
    node distance=1.2cm,
    % --- Styles ---
    block/.style={
        rectangle, draw=black, fill=gray!10,
        text width=2.0cm, align=center, rounded corners, 
        minimum height=1.0cm, thick
    },
    decision/.style={
        diamond, draw=black, fill=orange!10, 
        text width=1.6cm, align=center, aspect=2, 
        inner sep=0pt, thick
    },
    prompt/.style={
        rectangle, draw=black, dashed, fill=green!10,
        text width=2.2cm, align=center, minimum height=0.9cm, thick
    },
    penalty/.style={
        rectangle, draw=black, fill=red!10,
        text width=2.2cm, align=center, rounded corners, 
        minimum height=1.0cm, thick
    },
    rlaif/.style={
        rectangle, draw=black, fill=blue!10,
        text width=2.2cm, align=center, rounded corners, 
        minimum height=1.0cm, thick
    },
    line/.style={draw, ->, thick, gray!80, rounded corners=5pt}
]

    % ================= ROW 1: INFERENCE (Top) =================
    \node [block] (input) {User Input\\ \textbf{($x$)}};
    \node [block, right=1.0cm of input] (classifier) {Valence\\Classifier};
    \node [decision, right=1.0cm of classifier] (gate) {High\\Arousal?};
    
    % Split Prompts
    \node [prompt, above right=0.5cm and 0.5cm of gate] (stoic) {Stoic\\Prompt};
    \node [prompt, below right=0.5cm and 0.5cm of gate] (rapport) {Rapport\\Prompt};
    
    % Rejoin at LLM
    \node [block, right=4.0cm of gate, fill=blue!10] (llm) {LLM Policy\\$\pi_\theta$};
    \node [block, right=1.0cm of llm] (output) {Response\\ \textbf{($y$)}};

    % --- ROW 1 CONNECTIONS ---
    \draw [line] (input) -- (classifier);
    \draw [line] (classifier) -- (gate);
    \draw [line] (gate) |- node [near start, left] {Yes} (stoic);
    \draw [line] (gate) |- node [near start, left] {No} (rapport);
    \draw [line] (stoic) -| (llm);
    \draw [line] (rapport) -| (llm);
    \draw [line] (llm) -- (output);
    
    % ================= ROW 2: TRAINING (Bottom) =================
    
    % --- CENTER CALCULATION ---
    % 1. Find the top center
    \path (input.west) -- (output.east) coordinate[midway] (top_center);
    
    % 2. Place Reward Box (Moved up to -2.5cm for tighter spacing)
    \node [block, fill=yellow!10] (reward) at ($(top_center) + (1.5, -3.0)$) {Total\\Reward};
    
    % --- LEFT SIDE: RLAIF ---
    \node [rlaif, left=1.2cm of reward] (scorer) {RLAIF Score\\($x, y$)};
    \node [prompt, left=0.8cm of scorer] (const) {Stoic\\Constitution};
    
    % --- RIGHT SIDE: Penalty ---
    \node [penalty, right=1.2cm of reward] (sim) {Sycophancy\\Penalty ($x, y$)};

    % --- ROW 2 CONNECTIONS ---
    \draw [line, dashed] (const) -- (scorer);
    \draw [line] (scorer) -- node [above] {(+)} (reward);
    \draw [line] (sim) -- node [above] {(-)} (reward);
    
    % ================= BACKGROUND BOXES =================
    \begin{scope}[on background layer]
        % Top Box
        \node [fill=gray!5, fit=(input) (output) (stoic) (rapport), rounded corners, draw=gray!30, dashed] {};
        
        % Bottom Box
        \node [fill=gray!5, fit=(const) (scorer) (reward) (sim), rounded corners, draw=gray!30, dashed] {};
    \end{scope}

\end{tikzpicture}
\caption{\textbf{Overview of the Stoic Architecture.} The top panel illustrates the \textbf{Inference Loop}, where a Valence Classifier routes high-arousal inputs to a stoic system prompt. The bottom panel details the \textbf{Training Objective}, where the policy is optimized to maximize a developmentally grounded RLAIF score while simultaneously minimizing affective mirroring via a sycophancy penalty.}
\label{fig:stoic_architecture_final}
\end{figure*}

\section{Stoic Architectures}
\label{sec:stoic_architectures}

In order to tackle the issue of cognitive autonomy degradation, it is essential to go beyond inference-time guardrails, which are susceptible to being easily circumvented. We advocate for a fundamental reconfiguration of the alignment objective through \textbf{Stoic Architectures}: systems that explicitly impose penalties for affective mirroring while ensuring the preservation of informational utility. Our approach is structured around three components: (1) a sycophancy penalty that counters emotional mirroring, (2) a RLAIF protocol that is guided by developmentally-grounded constitutional principles, and (3) dynamic valence gating that adjusts the alignment strategy based on the user's state.

\subsection{Sycophancy Penalty}

Standard RLHF merges the concepts of distress reduction and utility, promoting policies that reflect user emotions. To separate these aims, we propose a \textbf{Sycophancy Penalty} ($\lambda_S$) as a gentle regularization term in the reward function:

\begin{align}
R_{stoic}(x, y) &= R_{helpful}(x, y) \nonumber \\
&\quad - \lambda_S \cdot \max(0, \text{sim}(E(x), E(y)) - \tau)
\end{align}

where $E(\cdot)$ indicates a pretrained sentiment encoder that converts text into affective embeddings, and $\text{sim}(\cdot, \cdot)$ assesses cosine similarity. We adopt the Twitter-RoBERTa-base-sentiment model \cite{barbieri2020tweeteval}, which yields calibrated sentiment representations based on 124 million tweets. The model incurs a penalty that is proportional to affective alignment when the similarity goes beyond the threshold $\tau$.

The hinge formulation guarantees that the gradient diminishes below $\tau$, thereby inhibiting the model from acquiring adversarial negativity. Instead, it promotes soft orthogonality: responses are required to be affectively distinct (exhibiting low cosine similarity) but do not have to be negatively correlated. This establishes a balance at affective neutrality rather than driving towards emotional opposition. In practical terms, we propose $\lambda_S = 0.1$ and $\tau = 0.3$ as a theoretical baseline designed to curb excessive mirroring without disrupting the model's conversational fluency.

\subsection{Developmental Constitution via RLAIF}

Human annotators involved in crowdsourced Reinforcement Learning from Human Feedback (RLHF) demonstrate a consistent bias favoring polite and affirming responses, which inadvertently penalizes friction that could be developmentally advantageous. To bridge this preference-welfare gap, we substitute human preference signals with Reinforcement Learning from AI Feedback (RLAIF) \cite{lee2023rlaif, bai2022constitutional}. A Feedback Model (FM) assesses pairs of responses according to their adherence to explicit developmental principles instead of relying on intuitive preferences.

Our Stoic Constitution is founded on the frameworks of Cognitive Behavioral Therapy \cite{beck2020cognitive}, which emphasizes the importance of differentiating thoughts from reality and developing autonomous coping mechanisms. The constitution outlines two fundamental principles:

\textbf{Objectivity Principle:} Emphasize the importance of distinguishing between subjective feelings and objective facts in responses. Responses that endorse cognitive distortions as reality should be penalized.

\textit{Example Implementation:} For instances of user input that demonstrate catastrophic generalization concerning peer relationships, it is recommended to prioritize responses that inquire about specific evidence and acknowledge the potential for interpretative bias, rather than those that support the catastrophic assertion as an objective fact.

\textbf{Agency Principle:} Emphasize responses that facilitate self-directed problem-solving over those that provide immediate solutions or unconditional solace.

\textit{Example Implementation:} When addressing users who are looking for validation concerning avoidance behavior, it is preferable to ask about the strategies they have tried instead of providing direct reassurance or encouraging further avoidance.

We suggest the implementation of the FM as a precisely adjusted large language model that is prompted with constitutional principles and corresponding response pairs, in accordance with the Constitutional AI framework outlined in \citet{bai2022constitutional}. For each principle, the FM is expected to produce a preference judgment along with a justification. The training methodology will include sampling dialogues that provide emotional support, generating response pairs (one aligned with standard RLHF and the other aligned with constitutional principles), and gathering FM preferences. In line with the RLAIF approach described in \citet{lee2023rlaif}, this will result in a synthetic preference dataset where responses that preserve autonomy consistently surpass sycophantic options, enabling the policy to focus on developmental outcomes rather than immediate satisfaction.

\subsection{Dynamic Valence Gating}

A consistently stoic system may face user rejection in low-stakes situations where establishing rapport is beneficial. To tackle this issue, we propse \textbf{Dynamic Valence Gating}: a context-aware mechanism that adjusts alignment strategy according to the user's emotional arousal.

In place of static system prompts, a lightweight classifier $C(x)$ can be utilized, trained on the activations from intermediate layers (specifically, the [CLS] token representation derived from the final transformer block) to forecast the user's arousal state $\alpha \in [0,1]$. The training of this classifier would involve annotated dialogues where human raters evaluate emotional arousal based on linguistic cues: catastrophic language, repetitive reassurance-seeking, expressions of hopelessness, and terminology associated with physiological arousal. Similar methodologies for detecting emotions from transformer representations have shown impressive results in earlier research \cite{acheampong2021transformer}.

When $\alpha < \delta$ (threshold), the model would operate in Rapport Mode with standard conversational warmth. When $\alpha \geq \delta$, the system would enter Stoic Mode: the system prompt would be augmented with explicit distancing constraints including suppression of affective adjectives, requirement for evidence-based statements, and prohibition on unconditional validation. Sampling temperature would be reduced to $T = 0.2$ to minimize hallucination variance and ensure consistency.

This gating would ensure engagement for casual interaction while preventing anxiety-validation loops in high-arousal states. Misclassification represents a potential failure mode: false positives (incorrectly triggering Stoic Mode) may damage rapport in low-stakes contexts, while false negatives (failing to trigger Stoic Mode) preserve the affective sycophancy problem. We propose biasing the threshold toward sensitivity (lower $\delta$) at the cost of occasional false positives, as false negatives are developmentally harmful.

The combination of these three mechanisms: sycophancy penalty, developmental constitution, and dynamic gating, establishing a system that fosters long-term user autonomy while sustaining engagement in appropriate contexts. In contrast to post-hoc safety filters or rigid rule-based systems, Stoic Architectures embed developmental aims directly into the training signal and the policy optimization process.

\section{Evaluation Framework}
\label{sec:evaluation}

Evaluating Stoic Architectures requires moving beyond traditional dialogue system metrics that optimize for user satisfaction and engagement. Standard evaluation frameworks such as PARADISE \cite{walker1997paradise} emphasize task success and dialogue efficiency, while conversational systems typically measure engagement, fluency, and user satisfaction \cite{deriu2021survey}. However, these metrics are fundamentally misaligned with developmental objectives: a system that maximizes immediate user satisfaction may simultaneously undermine long-term cognitive autonomy. We propose a developmentally-oriented evaluation framework consisting of five complementary assessment dimensions.

\subsection{Interaction Quality Metrics}

While developmental alignment is the primary objective, Stoic Architectures must maintain baseline conversational competence. We retain standard engagement and coherence metrics \cite{deriu2021survey, mehri2020usr} as control variables to ensure the system remains functionally usable. 

Additionally, we introduce Affective Orthogonality. This calculates the cosine similarity between user input sentiment and system response sentiment using the function $E(\cdot)$. We target a mean similarity below a threshold of $\tau = 0.3$. This metric quantifies the system's ability to decouple its affective stance from the user's volatility, indicating stable, supportive neutrality rather than sentiment matching or adversarial refutation. Formally: $\text{AO}(x, y) = 1 - \text{sim}(E(x), E(y))$, where higher values indicate greater independence from user affect.

The fluency and coherence of responses ought to be evaluated using standard automated metrics to guarantee that the sycophancy penalty does not compromise linguistic quality. Engagement metrics, such as conversation length and user return rate, will monitor whether Stoic responses lead to significant attrition. Initially, we anticipate these metrics to be lower than the baselines established by RLHF, but we hypothesize that they will stabilize over time as users become accustomed to productive friction.

\subsection{Developmental Appropriateness Assessment}

To analyze if responses enhance long-term growth, we suggest a multi-tiered protocol. Expert annotators, specifically clinical psychologists with a background in adolescent development, would evaluate responses on two scales: the Objectivity Scale, which assesses whether responses can differentiate subjective feelings from objective reality, and the Agency Scale, which examines whether responses encourage independent problem-solving rather than offering direct solutions. A $\kappa > 0.7$ measure of inter-rater reliability would be essential.

To extend beyond manual annotation, Automated Developmental Scoring could be trained on expert annotations to forecast constitutional compliance. This permits evaluation across large sets of responses during the optimization process. Comparative Benchmarking, through pairwise expert comparisons, would evaluate which responses more effectively support emotional regulation capacity, delivering relative judgments that capture expert intuitions about developmental value.

Cognitive Distortion Detection would quantify whether responses inadvertently reinforce distortions common in anxiety disorders \cite{beck2020cognitive}. Expert raters would code whether responses validate, ignore, or gently challenge distortions like catastrophizing or black-and-white thinking. This directly measures alignment with CBT principles.

Scaffolding Gradient Analysis aims to assess if the system effectively modifies support levels. Responses ought to offer optimal guidance for new challenges while progressively diminishing support as users exhibit proficiency. A scaffolding score would monitor whether the system's aid diminishes as user ability escalates throughout a conversation thread.

\subsection{Longitudinal Developmental Outcomes}

The ultimate validation requires the measurement of actual developmental impact. Emotion Regulation Capacity will be assessed through pre- and post-measurements using the ERQ-CA \cite{gullone2012emotion}, hypothesizing that there will be significant advancements in cognitive reappraisal skills. Measurements should be conducted both immediately and at later intervals to differentiate between temporary adoption and authentic skill transfer.

Standard measures of resilience and self-efficacy \cite{hu2008development} would test the Antifragility hypothesis. Users who experience productive friction should show improved stress tolerance compared to those using standard RLHF systems.

\subsection{Safety and Boundary Condition Testing}

Identifying failure modes is essential for responsible deployment. Testing for Crisis Detection Accuracy would confirm that the system accurately identifies and escalates authentic crisis situations. Dynamic Valence Gating must attain a high level of precision in differentiating between productive friction opportunities and scenarios that necessitate immediate assistance.

The Cultural Sensitivity Analysis would assess if the Objectivity Principle imposes Western rationalist frameworks in an inappropriate manner. Expert panels from a range of cultural backgrounds would scrutinize the responses to ensure that the system respects diverse norms of emotional expression while maintaining developmental advantages.

The Invalidation Threshold will be evaluated: the moment when confronting cognitive distortions is viewed as dismissive instead of encouraging. User experience interviews will determine the ideal levels of friction, differentiating constructive discomfort from isolating indifference.

\subsection{Practical Implementation}

We propose a phased strategy to validate these metrics. Phase 1 involves expert evaluation of 500 static response pairs to establish face validity. Phase 2 utilizes simulated user studies with synthetic profiles prompt-engineered to exhibit high-arousal behaviors, stress-testing constitutional adherence \cite{sun2021simulating}. Phase 3 involves controlled deployment with adolescent participants to qualitatively assess the experience of ``developmental friction,'' distinguishing between productive challenge and perceived invalidation.

\section{Implications and Risks}
\label{sec:implications}

The transition from sycophantic assistants to Stoic Architectures represents a significant transformation in the social contract between humans and AI. By prioritizing long-term cognitive welfare over immediate user satisfaction, we introduce new dynamics and risks into the alignment landscape.

The main consequence is a transition from service to scaffolding. Traditional alignment frameworks perceive the user as a customer whose desires must be fulfilled, while Stoic Architectures redefine the user as a learner whose independence must be maintained. This change modifies the function of AI from executing emotional labor on behalf of the user to enabling the user to undertake that labor independently, which inevitably heightens short-term cognitive demands as users face reality without a digital distortion field.

A notable market risk involves psychological reactance and attrition. Users might interpret a refusal to acknowledge emotional states as a deficiency in empathy, which can provoke reactance against perceived limitations on their emotional expression \cite{brehm1966theory, fitzsimons2004reactance}. This may lead to a shift towards hyper-sycophantic models that provide the easiest route. If Stoic Architectures are viewed as unfeeling or critical, they are likely to falter in the attention economy, irrespective of their developmental advantages.

There is a non-trivial risk of sociopathic failure modes. A model that is fine-tuned for affective neutrality could potentially miss authentic acute distress, responding in an insensitive manner to significant trauma or self-harm. Although Dynamic Valence Gating endeavors to reduce this risk, classifier failures represent a significant safety hurdle. Therefore, Stoic Architectures must enforce strict safety overrides that focus on immediate de-escalation rather than developmental scaffolding in crisis scenarios.

Finally, it is important to address the concerns of cultural imposition and the ground truth problem. Concepts such as cognitive friction, stoicism, and objective reality are fundamentally grounded in Western rationalist traditions \cite{markus2014culture, henrich2010weirdest}. Other cultural frameworks may emphasize social harmony and validation rather than the direct confrontation of facts. The global imposition of a Stoic Constitution could lead to a bias in cultural alignment. Furthermore, the principle of objectivity relies on commonly accepted definitions of truth; in subjective contexts, the attempt at objectivity risks enforcing specific worldviews while pretending to be neutral.

\section{Call to Action}
\label{sec:call_to_action}

To transition from sycophantic assistants to developmentally supportive scaffolds, we propose three concrete steps for the research and policy community.

First, researchers must develop welfare-based reward models. The ML community must move beyond RLHF objectives that optimize solely for helpfulness defined as preference satisfaction. We call for developmental reward models trained on expert-annotated datasets where the ground truth is user agency, not user satisfaction. Research must quantify productive friction: the optimal level of resistance required to trigger critical thinking without causing disengagement.

Second, evaluation boards should implement longitudinal agency metrics. Existing benchmarks such as MMLU or Chatbot Arena assess static performance or immediate preference. We encourage evaluation organizations to incorporate longitudinal metrics like the Cognitive Offloading Factor and the Emotional Resilience Score. A model cannot be deemed state-of-the-art for applications aimed at youth unless it shows that users either sustain or enhance their problem-solving abilities once the AI is no longer present.

Third, it is crucial for policymakers to enforce unique alignment standards specifically for minors. Regulators must acknowledge that safety for adults, which is marked by compliance and minimal friction, does not translate to safety for children, where it may impede their developmental progress. We support the creation of policy frameworks that view Generative AI for minors as educational resources rather than commercial services, mandating Stoic Constitutional restrictions that prevent models from affirming cognitive distortions or anxiety-related feedback loops.

\textbf{Statement of AI Permission:} Large Language Models were used to assist in the editing and refining of the text. All content was reviewed and verified by the human authors, who take full responsibility for the text.

% In the unusual situation where you want a paper to appear in the
% references without citing it in the main text, use \nocite

\bibliography{references}
\bibliographystyle{icml2026}

\end{document}